\documentclass[a4paper]{jpconf}
\usepackage{graphicx}
\usepackage{subfigure}
\usepackage[fleqn]{amsmath}

\newcommand{\unit}[1]{\,\mathrm{#1}}
\newcommand{\ee}{\mathrm{e}}

\begin{document}
\title{Towards a new 
quark-nuclear matter EoS for applications in astrophysics and heavy-ion collisions}

\author{Niels-Uwe Bastian$^1$, David Blaschke$^{1,2,3}$}

\address{$^1$Institute for Theoretical Physics, University of Wroc{\l}aw, 50-204 Wroc{\l}aw, Poland}
\address{$^2$Bogoliubov Laboratory for Theoretical Physics, JINR Dubna,
141980 Dubna, Russia}
\address{$^3$National Research Nuclear University (MEPhI), 115409 Moscow, Russia}

\ead{niels-uwe@bastian.science}

\begin{abstract}
The aim of our work is to develop a unified equation of state (EoS) for nuclear and quark matter for a wide range in temperature, density and isospin so that it becomes applicable for heavy-ion collisions as well as for the astrophysics of neutron stars, their mergers  and supernova explosions.
As a first step, we use improved EoS for the hadronic and quark matter phases and join them via Maxwell construction. 
We discuss the limitations of a 2-phase description and outline steps beyond it, towards the formulation of a unified quark-nuclear matter EoS on a more fundamental level by a cluster virial expansion.
\end{abstract}

\section{Introduction}
The description of the hadron-to-quark matter phase transition (PT) in effective models for the equation of state (EoS) of strongly interacting matter is still a challenge, even half a century after the quark model 
has been developed. 
A modern description should consistently address the question how the breaking and restoration of the symmetries of the QCD Lagrangian take place as a function of the thermodynamic variables, temperature $T$ and baryon chemical potential $\mu_b$ (for asymmetric matter, there is a set of additional chemical potentials which eventually can be restricted by conditions like conservation laws).
Particular emphasis is hereby on the dynamical chiral symmetry breaking and restoration, an aspect that can be very effectively addressed within models of the Nambu--Jona-Lasinio (NJL) type with the quark condensate or the dynamically generated quark mass serving as order parameters of the transition. 
A crucial role is played by quark and gluon confinement, i.e. the empirical principle that these 
QCD degrees of freedom carrying color charges have to be confined within color neutral bound states in the hadronic phase. 
Accordingly, within the deconfinement PT quarks and gluons appear as quasiparticles in the system and together with the chiral symmetry restoration in the quark sector the hadronic bound states get dissociated (Mott effect).
However, as NJL models have no confinement, modifications are invoked to remedy this fact. Let us mention four possibilities: 
(i) a bag pressure $B$, (ii) a confining meanfield (density functional)
\cite{Ropke:1986qs,Khvorostukin:2006aw,Li:2015ida}, 
(iii) minimal coupling of the quark dynamics to the Polyakov-loop with an effective potential \cite{Ratti:2005jh}, and (iv) a nonlocal, covariant generalization of the current-current coupling \cite{Schmidt:1994di,Bowler:1994ir,GomezDumm:2005hy}. 
A benchmark check for any effective model is the comparison with EoS data from lattice QCD simulations limited at present to $T$ for vanishing $\mu_b$ \cite{Borsanyi:2013bia,Bazavov:2014pvz}.

The working horse of the QCD PT model studies is still the so-called 2-phase description, where separately developed EoS for the hadronic and the quark-gluon phases of matter are employed to  
obtain a hybrid quark-nuclear matter EoS by applying Gibbs conditions of phase equilibrium. See, e.g., Ref.~\cite{Barz:1989cv} for a phase diagram.

In this contribution, we present a hybrid EoS obtained with a 2-phase construction using modern inputs in order to illustrate deficiencies that are inherent in such an approach when applied to the QCD transition.

Finally, we outline steps beyond a 2-phase description towards the formulation of a unified quark-nuclear matter EoS on the basis of a cluster virial expansion for quark matter.
While being similar in spirit to developments for describing clusters in nuclear matter \cite{Ropke:2012qv}, for the present task one has to address additionally the questions of chiral symmetry restoration and deconfinement within a relativistic, field-theoretic approach. 
To this end, the $\Phi$-derivable approach \cite{Baym:1961zz} shall be generalized as 
described in \cite{Blaschke:2015bxa}.

\section{Hybrid EoS from 2-phase construction with modern inputs}
For the hadronic side we use the well known density-dependent relativistic meanfield theory parametrisation (DD2) by Stefan Typel \cite{Typel:1999yq,Typel:2009sy}.
This EOS reproduces very well the known behavior of nuclear matter near the saturation density for symmetric matter and describes a liquid-gas PT.
For applications at high temperatures ($T>50\unit{MeV}$) the hadronic side should be improved by taking into account the appearance of additional hadrons (Pions, Kaons\dots).

\begin{figure}[htbp] 
\begin{minipage}[c]{0.73\textwidth}
    \includegraphics[scale=0.44]{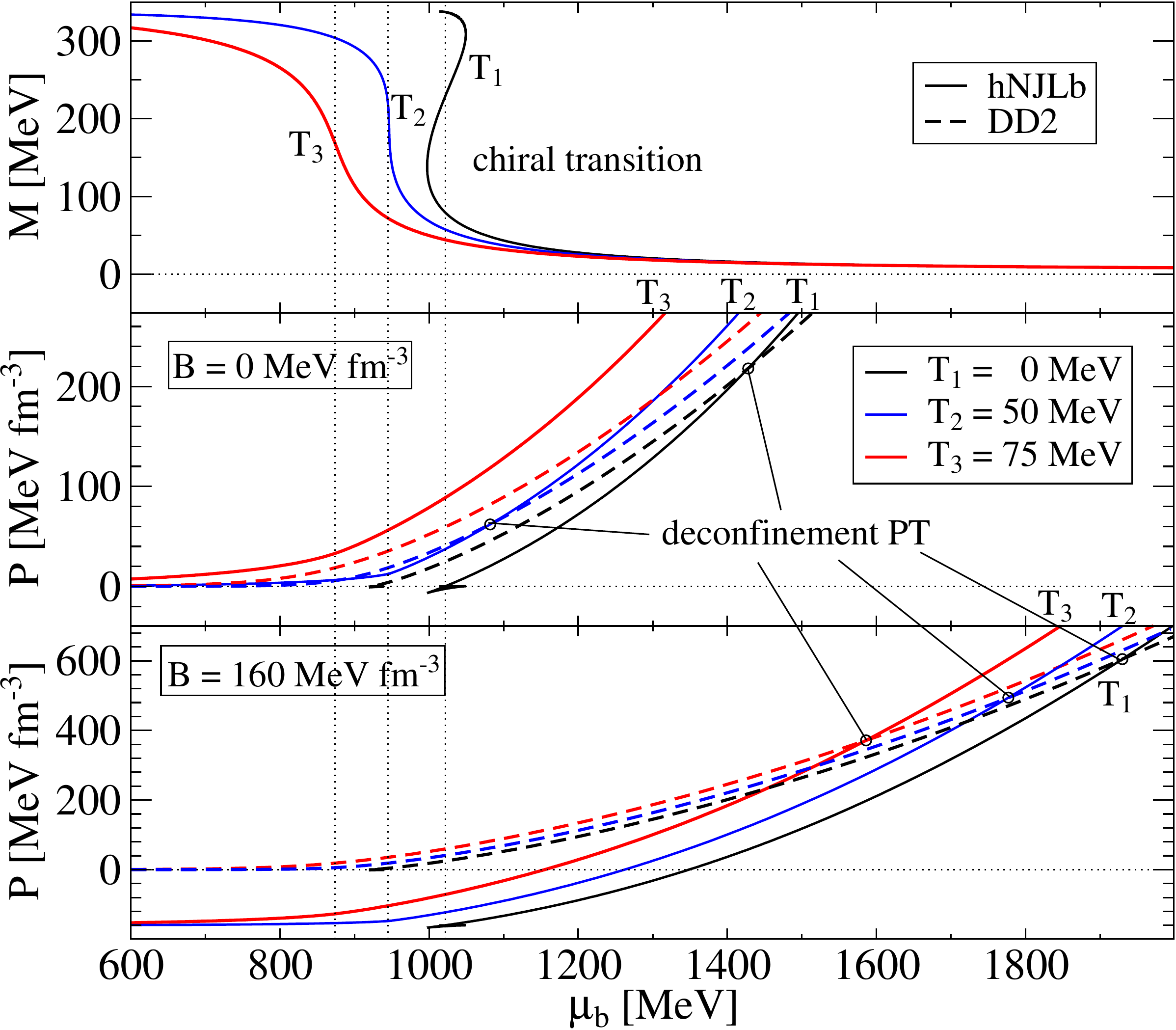}
\end{minipage}\hfill
\begin{minipage}[h]{0.27\textwidth}
    \caption{Effective mass $M$ (upper panel) and pressure $P$ (middle and lower panels) as functions of the baryochemical potential $\mu_b$ for the hNJLb EoS for $\eta_{02} = 0.08$ and $\eta_ {04} = 5.0$ in the isospin symmetric case. 
    The chiral transition is defined by the hNJLb EoS and deconfinement occurs at the crossing points with the hadronic (DD2) EoS for $B=0$ (middle panel) and for $B=160$ MeV/fm$^{3}$ (lower panel).}
    \label{fig:pt_construction}
\end{minipage}
\end{figure}

\begin{figure}[!htb] 
\includegraphics[scale=0.44]{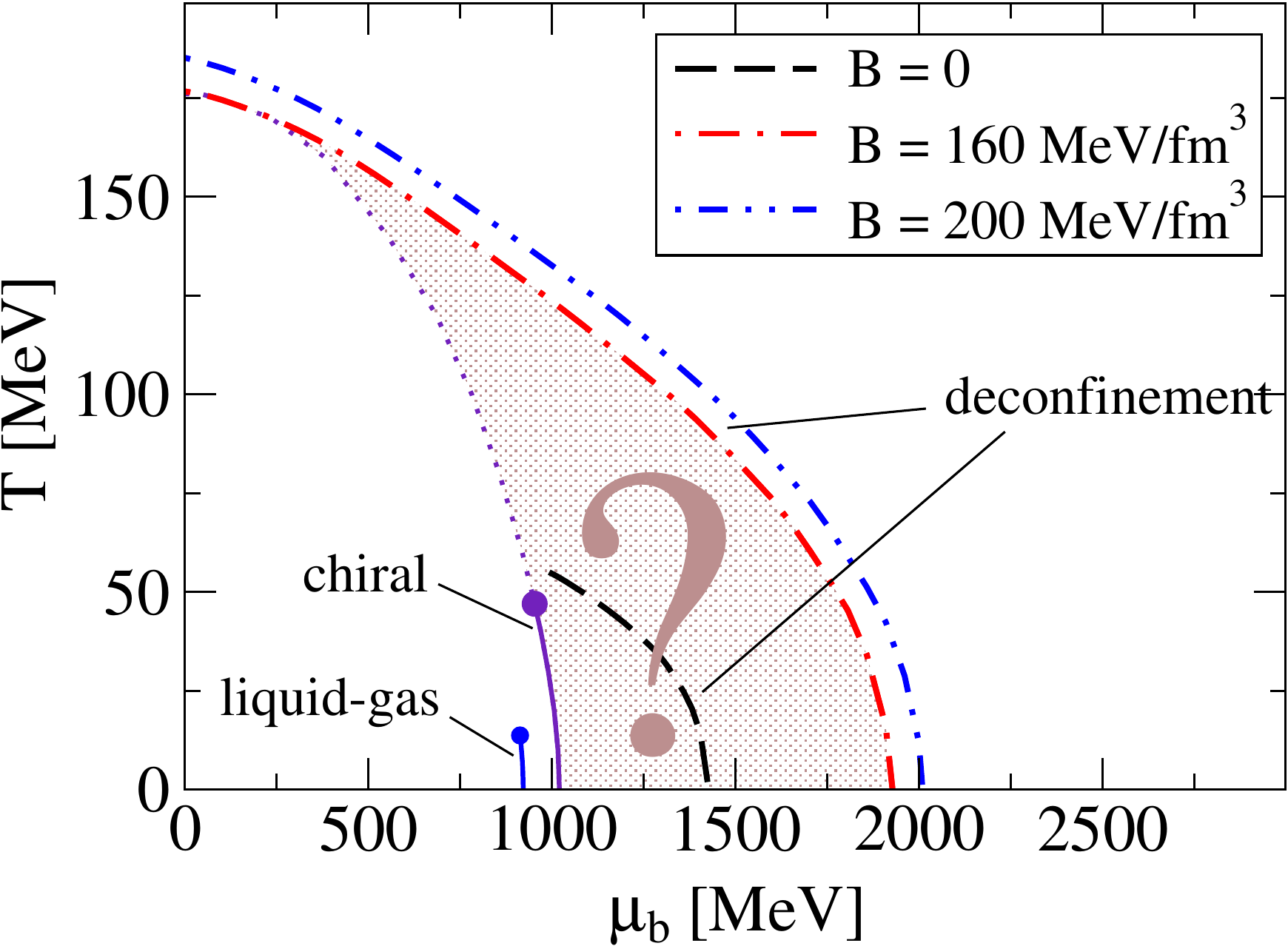}  \hfill
\includegraphics[scale=0.44]{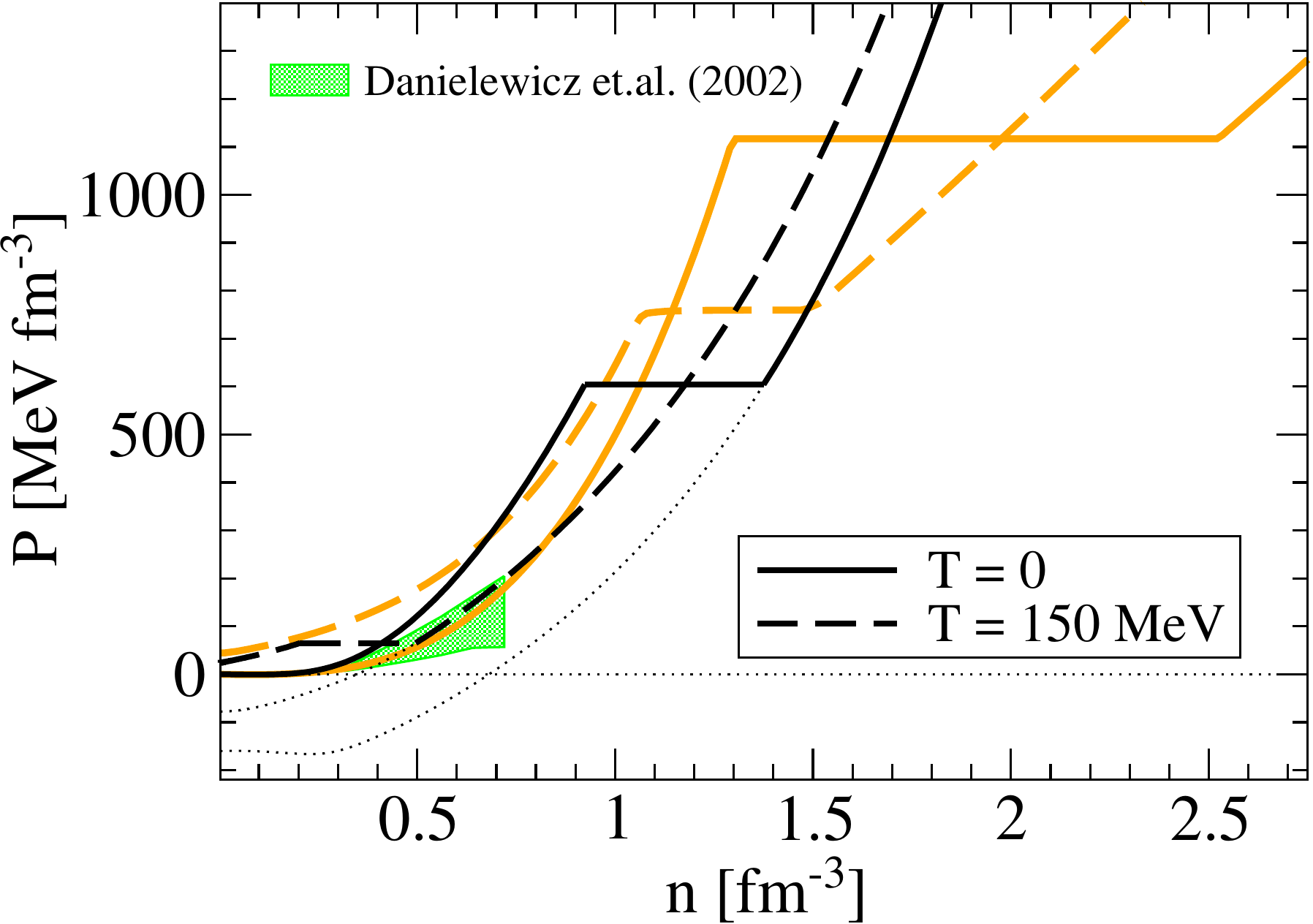}
\caption{Left: Phase diagram for isospin symmetric quark-nuclear matter.
The nuclear liquid-gas transition (blue line) and the chiral symmetry restoration (purple line) are first order PTs with endpoints. The deconfinement transition between DD2 and hNJLb is shown for 
different bag pressures: $B=0$ (dashed line), $B=160$ MeV/fm$^3$ (dash-dotted line) and $B=200$ MeV/fm$^3$ (dash-double-dotted line).
Right: EoS with deconfinement transition between DD2 and hNJLb for $\eta_{02} = 0.08$, $\eta_ {04} = 5.0$ and bag constant 
$B = 160 \unit{MeV/fm^{3}}$ compared to the EoS from Ref.~\cite{Khvorostukin:2006aw} which is used in three-fluid hydro simulations \cite{Ivanov:2010cu,Ivanov:2015vna}.
Shown are isothermes for $T=0$ (solid lines) and $T=150$ MeV (dashed lines) in symmetric matter and the flow constraint of Ref.~\cite{Danielewicz:2002pu}}
\label{fig:pt_povern}
\end{figure}

For the quark side we use the higher order NJL (hNJL) model recently developed by Sanjin Beni\'c for zero temperature \cite{Benic:2014iaa} and subsequently applied to hybrid compact stars \cite{Benic:2014jia}.
For extending it to finite temperature, we use the grand potential density
(symmetric case)
\begin{align}
	\Omega(T,\mu) &= U(\sigma,n) - 2N_c N_f T\int \frac{\mathrm d^3p}{(2\pi)^3} \left\{ 
	\frac{\tilde{E}}{T} + \ln\left\{\left[1 + \ee^{-\frac{(\tilde{E} - \tilde\mu)}{T}}\right]\left[1 + \ee^{-\frac{(\tilde{E} + \tilde\mu)}{T}}\right]\right\}\right\} + \Omega_0
\end{align}
with an effective potential $U(\sigma, n)$ and a renormalized quark chemical potential $\tilde{\mu}=\mu-\Delta_\mu(n)$. The renormalized mass $M=m-\Delta_m(\sigma)$ in the dispersion relation $\tilde{E}=\sqrt{p^2+M^2}$ 
is obtained from solving the gap equation, i.e. minimizing $\Omega(T,\mu)$ \cite{Benic:2014iaa,Benic:2014jia} and thus defining the quark matter pressure in equilibrium $P_\mathrm{hNJL}(T,\mu)=-\Omega(T,\mu)$.
In Fig.~\ref{fig:pt_construction} we show the solution of the mass gap equation (upper panel) and the corresponding quark pressure (middle panel) as a function of $\mu_b=3\mu$ for selected isothermes which shows the 
chiral PT as a van-der-Waals wiggle similar to that for the nuclear liquid-gas transition in the DD2 EoS. Both EoS have a critical endpoint with a temperature above which the transition is a crossover, see
Fig.~\ref{fig:pt_povern}.
The crossing points with the hadronic pressure are denoted as {\it deconfinement PT} when for increasing $\mu_b$ the quark pressure lies above the hadronic one and is thus preferred.
We notice that at finite $T$ unphysical second crossings occur and that already above $T = 55$ MeV no hadronic matter exists. This unphysical behaviour is due to the absence of confinement in the hNJL model.
We cure this pathology in the simplest way, by including a bag pressure
$	P_\mathrm{hNJLb} = P_\mathrm{hNJL} - B$~,
see Fig.~\ref{fig:pt_construction} (lower panel) for two values of $B$
that result in PT lines in the phase diagram of Fig.~\ref{fig:pt_povern}
(left) that correspond to critical temperatures in the range $175 - 180$
MeV at $\mu_b=0$.

The left panel of Figure \ref{fig:pt_povern} shows the phase transition between DD2 and hNJLb with $B = 160 \unit{MeV fm^{-3}}$ for two temperatures $T = 0$ and $T = 150\unit{MeV}$ in the pressure over density plane compared to the EoS used by Yuri Ivanov in three-fluid hydrodynamic simulations \cite{Ivanov:2010cu,Ivanov:2015vna}.
For our EoS, the deconfinement PT  occurs at much lower densities, in particular for the temperatures expected in heavy-ion collisions for NICA and FAIR energies, what should result in interesting modifications of the baryon stopping signal of this transition predicted in \cite{Ivanov:2010cu}.


\section{Discussion and outlook}

Applying the 2-phase construction as described above reveals a crucial problem of such an approach to the quark-nuclear transition.
At lower chemical potentials a chiral transition occurs within the NJL model, so we have chiral restoration of the quarks and they lose their masses.
But in the construction at this point there are no quarks and only the hadrons exist, for which no chiral transition is considered.
Therefore, in the resulting EOS the chiral transition is moved to the deconfinement transition, by the PT construction.
On the other hand, if in the hatched region of Fig.~\ref{fig:pt_povern} the quarks should already lose their masses while being still confined in hadrons, this should alter the hadron properties (e.g., the radii) and thus modify the behavior of their EoS in crucial way, so that the deconfinement transition probably should appear much earlier than shown in that figure, but this will probably still leave a window for a so-called {\it quarkyonic phase}
\cite{McLerran:2007qj} which may end in a triple point \cite{Andronic:2009gj}.
Such a phase might not exist and it can be avoided by construction,
defining a temperature dependent bag constant so that the deconfinement phase transition coincides with the chiral phase transition \cite{Klahn:2015mfa}.
On the other hand, one can explore this window between defined asymptotic
EoS of hadronic and quark matter by interpolation 
\cite{Masuda:2012ed,Kojo:2014rca,Kojo:2015fua}, thus admitting the lack 
of fundamental theoretical concepts in that region.

To develop a unified quark-nuclear EoS we suggest to work out a cluster expansion \cite{Blaschke:2015bxa} on the basis of the $\Phi$-derivable approach \cite{Baym:1961zz}, where hadrons are considered as bound states of quarks. 
Their dissociation in hot, dense matter due to the Mott effect will be triggered by the lowering of the quark continuum thresholds that inevitably renders them unbound, but still allows their presence as strong correlations in the scattering continuum. This situation is accounted for within a generalized Beth-Uhlenbeck EoS that follows within this approach. 
As opposed to NJL models which have no confinement mechanism (and also their Polyakov-loop generalization which becomes ambigious at finite chemical potentials), we propose here the usage of relativistic density functionals which effectively account for confinement and can be motivated, e.g., within the concept of string-flip model \cite{Ropke:1986qs}. 
Then, in the confined, hadronic phase before the Mott transition quark exchange effects among hadrons, due to their wave function overlap, will give rise to strongly repulsive forces. These Pauli blocking effects in dense hadronic matter justify the adoption of excluded volume prescriptions in the hadronic EoS in order to effectively account for them, see 
\cite{Benic:2014jia} for a recent application.
In this way one should obtain a consistent unified description of quark-nuclear matter in the QCD phase diagram.

\section*{Acknowledgement}
This work was supported by NCN under contract number 
UMO-2011/02/A/ST2/00306.

\section*{References}
\providecommand{\newblock}{}

\end{document}